\documentclass[11pt,preprint]{aastex}
\newcommand{\be}{\begin{equation}}
\newcommand{\ba}{\begin{eqnarray}}
\newcommand{\ee}{\end{equation}}
\newcommand{\ea}{\end{eqnarray}}
\newcommand{\etal}{et al.\ }
\newcommand{\etalb}{et al.}
\newcommand{\nb}{\bar{n}}
\newcommand{\Ni}{N_{\rm ion}}
\newcommand{\Omm}{\Omega_m}
\newcommand{\HI}{\rm H\,I}

\begin{document}
\title{Did the Universe Reionize at Redshift Six?}

\author{Rennan Barkana}
\affil{Canadian Institute for Theoretical Astrophysics, 60 St. George 
Street, Toronto, Ontario, M5S 3H8, CANADA; \\{\rm Present address:}\,
School of Physics and Astronomy, Tel Aviv University, Tel Aviv 69978,
ISRAEL}
\email{barkana@wise.tau.ac.il}

\begin{abstract}

In light of recent observations of spectra of the quasars SDSS
1030+0524 \citep{z6.3} and SDSS 1044-0125 \citep{z5.8}, we study the
observational signatures of different stages of the reionization
epoch. During the initial, pre-overlap stage, the hydrogen throughout
the universe is neutral except for isolated \ion{H}{2} bubbles due to
individual ionizing sources. The central stage of reionization is the
overlap stage, during which the individual \ion{H}{2} bubbles overlap
each other and reionize the low-density gas which takes up most of the
volume of the universe. Some neutral hydrogen remains in dense clumps
which are then slowly reionized during post-overlap. We show that both
of the recent observations are consistent with the post-overlap
stage. \citet{z6.3} may have observed the universe in the era before
the end of overlap, but a conclusive proof of this requires the
observation of similar intervals of Gunn-Peterson absorption at the
same redshift along several additional lines of sight.

\end{abstract}

Key Words: galaxies: high-redshift, cosmology: theory, 
galaxies: formation

\vspace{-.1in}
PACS: 98.80.-k, 98.65.Dx, 98.62.Ai

\section{Introduction}

\label{sec:Intro}

Until recently, the highest redshift detected for a quasar was $z=5.8$
\citep{f00}. Although the spectrum showed strong absorption
short-ward of the Lyman-$\alpha$ line of the quasar, the detection of
transmitted flux in this spectrum was taken to imply that reionization
was complete by $z \sim 6$, since even a small neutral fraction in the
IGM would have produced total absorption, a fact first pointed out by
\citet{GP}. The most natural explanation for reionization is 
photo-ionizing radiation produced by an early generation of stars and
quasars; recent calculations of structure formation in cold dark
matter (CDM) models find that reionization should naturally occur at
$z\sim 6$--12 \citep[and for comprehensive reviews on the subject of
reionization and the first galaxies, see \citealt{me01,me01b}]{hl97,
hl98, go97, chiu, g00}.

The reionization of hydrogen is expected to involve several distinct
stages. The initial, ``pre-overlap'' stage (using the terminology of
\citealt{g00}) consists of individual ionizing sources turning on and
ionizing their surroundings. The radiation from the first galaxies
must make its way through the surrounding gas inside the host halo,
then through the high-density region which typically surrounds each
halo. Once they emerge, the ionization fronts propagate more easily
into the low-density voids, leaving behind pockets of neutral,
high-density gas. During this period the IGM is a two-phase medium
characterized by highly-ionized regions separated from neutral regions
by sharp ionization fronts. The central, relatively rapid ``overlap''
phase of reionization begins when neighboring \ion{H}{2} regions start
to overlap. Whenever two ionized bubbles are joined, each point inside
their common boundary becomes exposed to ionizing photons from both
sources. Therefore, the ionizing intensity inside \ion{H}{2} regions
rises rapidly, allowing those regions to expand into high-density
gas. This process leads to a state in which the low-density IGM is
highly ionized and ionizing radiation reaches everywhere except for
gas located inside self-shielded, high-density clouds. Some neutral
gas does, therefore, remain in high-density structures which
correspond to Lyman Limit systems and damped Ly$\alpha$ systems seen
in absorption at lower redshifts. The ionizing intensity continues to
grow during this ``post-overlap'' phase, as an increasing number of
ionizing sources becomes visible to every point in the IGM.

Recently, \citet{z5.8} and \citet{z6.3} found, for the first time,
broad regions of Gunn-Peterson absorption in quasar spectra, with high
optical depth throughout each region. \citet{z5.8} observed the quasar
discovered by \citet{f00} with a higher spectral resolution and high
signal to noise, and discovered a dark region which is $\sim 5$ Mpc
long at $z \sim 5.3$. \citet{z6.3} discovered the most distant source
of light observed to date, a quasar at $z=6.28$; the spectrum showed a
dark region of $\sim 15$ Mpc length extending right up to $z = 6.15$,
where the emission of the quasar itself may begin to have an
effect. Since even a small neutral fraction in the IGM can produce
these long stretches of strong absorption, it has thus far been
unclear whether these observations indeed herald the empirical
discovery of the overlap era; instead, these observations may only be
approaching the end of overlap, which itself occurred at an even
higher redshift than that probed by current observations.

In this paper, we construct a model of the statistics of neutral and
ionized regions during reionization, in order to interpret the recent
observations. Our goal is to use the observations to infer the state
of the IGM at $z \sim 6$, and specifically to determine whether the
universe has indeed been observed during or even before the overlap
stage. We focus on the observable property which is not found at low
redshifts but is a new feature of the recent observations, namely long
intervals of continuous strong Ly$\alpha$ absorption. The model must
be statistical, since the ionizing intensity in the IGM is extremely
inhomogeneous during reionization and for some period afterwards.
Given the small sample of lines of sight probed thus far, we seek to
address the question of whether a statistical fluctuation could
produce the observed Gunn-Peterson absorption even if the entire IGM
has already been reionized at the observed redshift. Since the
properties of ionizing sources at these high redshifts are speculative
at present, having almost no observational constraints, we strive to
obtain model-independent conclusions. We present our model in \S
\ref{sec:Model} and the results in \S \ref{sec:Results}. We then
confront the model results with the observations in \S
\ref{sec:Disc}. Finally, we give our conclusions in \S \ref{sec:Conc}.

The basic theoretical framework in which the matter content of the
universe is dominated by CDM has recently received a major
confirmation from measurements of the cosmic microwave background
\citep{Boomerang,Maxima,Dasy}. Based primarily on these measurements, 
in this paper we use cosmological parameters $\Omm$ = 0.35,
$\Omega_\Lambda$ = 0.65,$\Omega_b = 0.05$, $\sigma_8 = 0.85$, $n=1$,
and $h=0.65$, where $\Omm$, $\Omega_\Lambda$, and $\Omega_b$ are the
total matter, vacuum, and baryon densities in units of the critical
density, $\sigma_8$ is the root-mean-square amplitude of mass
fluctuations in spheres of radius $8\ h^{-1}$ Mpc, and $n=1$
corresponds to a primordial scale-invariant power spectrum.

\section{Modeling the Statistics of Gunn-Peterson absorption}

\label{sec:Model}

\subsection{Ionizing Sources}

\label{sec:SourceModel}

In this section we describe a model which allows us to predict the
observable properties of quasar spectra during the various stages of
reionization. Roughly speaking, the question we wish to address is the
following; if we consider spectra taken along random lines of sight
through the universe, at a given redshift, then what fraction of these
spectra will be similar to those that have actually been observed? We
define the similarity of spectra in terms of the presence or absence
of long intervals of strong Ly$\alpha$ absorption. The absorption
depends on the distribution of gas density in the IGM and on the
ionization state of the gas, which in turn depends on the properties
of ionizing sources. Our strategy is to construct a model which is
general enough to include a wide range of possible scenarios, given
the current paucity of direct observational constraints on, for
example, the properties of the ionizing sources. While we study the
dependence of the results on the various input parameters, we
especially seek a model-independent probe of the reionization state of
the IGM. In this subsection we describe our model for the properties
of ionizing sources; in \S \ref{sec:IGMModel} we model the gas density
and ionization state, and obtain its opacity; finally, in \S
\ref{sec:FinModel} we discuss the physical quantities that we
calculate in order to compare our model with the observations.

Galaxies and the ionizing sources within them form in halos in which
gas can accumulate and cool. At high redshift, gas can cool
efficiently in halos down to a virial temperature of $\sim 10^4$ K or
a circular velocity of $V_c\sim 16.5\ {\rm km\ s}^{-1}$ with atomic
cooling. Cooling is possible down to $ V_c\sim 2\ {\rm km\ s}^{-1} $
with molecular hydrogen (H$_2$) cooling, but molecular hydrogen is
fragile and is expected to be photo-dissociated well before
reionization unless mini-quasars\footnote{A mini-quasar is an
accreting central black hole in a dwarf galaxy.} make a large
contribution to the ionizing intensity \citep{hrl97,har00}. Before
reionization, the IGM is cold and neutral, and these cooling
requirements set the minimum mass for halos which can host
galaxies. During reionization, however, when a volume of the IGM is
ionized by stars, the gas is heated to a temperature $T_{\rm IGM}\sim
1$--$2 \times 10^4$ K. We adopt a standard temperature of $T_{\rm
IGM}=1.5 \times 10^4$ K, and then the linear Jeans mass corresponds to
a virialized halo with a circular velocity of \be V_J=82
\left(\frac{T_{\rm IGM}} {1.5\times 10^4 {\rm K}}\right)^{1/2}\ {\rm
km\ s}^{-1}\ , \ee where this value is essentially independent of
redshift. Even halos well below the Jeans mass can pull in some gas
once the dark matter collapses to the virial overdensity. For
simplicity, we adopt a sharp cutoff associated with this suppression,
at a circular velocity of $V_c = V_J/2$, based on the results of
numerical simulations
\citep{tw96,qke96,whk97,ns97,ki00}. However, this pressure suppression
is not expected to cause an immediate suppression of the cosmic star
formation rate, since even after fresh gas infall is halted the gas
already in galaxies continues to produce stars, and mergers among
already-formed gas disks also trigger star formation. However, if the
overlap phase of reionization is relatively gradual then there may be
time for the suppression of star formation to become significant even
before the end of overlap.

Once gas collects inside a halo and cools, it can collapse to high
densities and form stars or a mini-quasar. Regardless of the nature of
the source, its ability to form is determined by gas accretion which,
in a hierarchical model of structure formation, is driven by mergers
of dark matter halos. Therefore, in order to determine the lifetime of
a typical source, we first define the age of gas in a given halo using
the average rate of mergers which built up the halo.  Based on the
extended Press-Schechter formalism \citep{lc93}, for a halo of mass
$M$ at redshift $z$, the fraction of the halo mass which by some
higher redshift $z_2$ had already accumulated in halos with galaxies
is \be F_M(z,z_2) = {\rm Erfc} \left(\frac{1.69/D(z_2)-
1.69/D(z)}{\sqrt{2 (S(M_{\rm min}(z_2))-S(M))}} \right)\ , \ee where
$D(z)$ is the linear growth factor at redshift $z$, $S(M)$ is the
variance on mass scale $M$ (defined using the linearly-extrapolated
power spectrum at $z=0$), and $M_{\rm min}(z_2)$ is the minimum halo
mass for hosting a galaxy at $z_2$ (as determined by the cutoff $V_c$
which was discussed above). We define the total age of gas in the halo
as the time since redshift $z_2$ where $F_M(z,z_2) = 0.2$, so that
most ($80\%$) of the gas in the halo has formed stars only since
then. Thus, e.g., at $z=6$ the age of the universe is $9.1 \times
10^8$ while the age of gas in a $3 \times 10^8 M_{\sun}$ halo is
$3.6\times 10^8$ yr ($z_2=8.8$), and the age of gas in a $10^{11}
M_{\sun}$ halo is $5.5\times 10^8$ yr ($z_2=12.0$). Low-mass halos
form out of gas which has recently cooled for the first time, while
high-mass halos form out of gas which has already spent previous time
inside small galaxies. We emphasize that the age we have defined here
is not the formation age of the halo itself, but rather it is an
estimate for the total period during which the gas which is currently
in the halo participated in star formation. However, the rate of gas
infall is not constant, and even within the galaxy itself, the gas may
not form stars at a uniform rate. The details involve complex
astrophysics, so we account for the general possibility of bursting
sources by adding a parameter $\zeta$, the duty cycle. We pick a
standard value of $\zeta=0.25$, which implies that each source is four
times brighter during a lifetime which is four times shorter than the
gas infall age that we specified above. This value of $\zeta$ is
motivated by the fluctuating accretion rate seen in simulations, but
in our model we allow for a wide range of possible values of $\zeta$.

Consider an \ion{H}{2} region produced by a source residing in a halo
of total mass $M$ and baryon fraction $\Omega_b/\Omm$. Since
recombinations within the \ion{H}{2} bubble are relatively
unimportant, we assume that the radius of the bubble is close to the
maximum radius $r_{\rm max}$ that is simply determined by equating the
number of ionized hydrogen atoms to the total number of ionizing
photons produced by the source over its lifetime. We also assume for
simplicity that each bubble is near its maximum size; this is a safe
assumption for our purposes since the neutral fraction is relatively
high at the outskirts of each \ion{H}{2} bubble, and so the
cross-section for observing a region with low optical depth is
dominated by the central regions of the bubbles. To estimate the
radius, we assume that the baryons in the halo are incorporated into
stars with an efficiency of $f_{\rm star}=10\%$, and that the escape
fraction for the resulting ionizing radiation is $f_{\rm esc}=5\%$,
where this low escape fraction is understood to include the effect of
recombinations in the dense gas within the source galaxy, its host
halo, and the relatively dense immediate surroundings of the halo.  If
the stellar IMF is similar to the one measured locally
\citep{scalo} then $\approx 4000$ ionizing photons are produced per
baryon in stars (for a metallicity equal to $1/20$ of the solar value;
Haiman, personal communication). We define a parameter which gives the
overall number of ionizations per baryon,
\be \Ni \equiv 4000 \, f_{\rm star}\, f_{\rm esc}\ . \ee Denoting by
$\nb_H^0$ the present mean number density of hydrogen, the radius of
the \ion{H}{2} bubble is
\be r_{\rm max}= \frac{1}{1+z} \left(\frac{3}{4\pi}\, \frac{\Ni}
{\nb_H^0}\, \frac{\Omega_b}{\Omm}\, \frac{M}{m_p} \right)^{
\frac{1}{3}} = 76\, {\rm kpc} \left(\frac{7}{1+z} \right) \left( 
\frac{\Ni}{20}\, \frac{M} {10^9 M_{\sun}}\, \frac{0.148}{\Omm h^2}
\right)^{\frac{1}{3}}\ . \label{rmax} \ee Here and throughout this 
paper we use proper, not comoving, distances. Note that the radius
$r_{\rm max}$ is larger than the halo virial radius by a factor of
$\sim 20$ that is almost independent of redshift and halo mass. Also
note that we would obtain a similar result for the size of the
\ion{H}{2} region around a galaxy if we considered a mini-quasar
rather than stars. This results from the high expected escape fraction
for mini-quasars \citep[$\sim 50\%$, e.g.,][]{wl00} which, together
with the high radiative efficiency of $\sim 6\%$, overcomes the low
efficiency \citep[$\sim 0.2$--$0.6\%$, e.g.,][]{m98} for incorporating
the baryons in a galaxy into a central black hole.

\subsection{Opacity of the IGM}

\label{sec:IGMModel}

Before the end of overlap, some large voids in the IGM are still
neutral, and they completely block out the flux at the Ly$\alpha$
wavelength. However, even at a wavelength $\lambda$ which corresponds
to redshifted Ly$\alpha$ inside an \ion{H}{2} region, the optical
depth rises with redshift. There may in general be two sources of
opacity at high redshift. First, the gas inside the \ion{H}{2} region
absorbs light strongly in the area of the resonance of the Ly$\alpha$
line, and can produce a significant opacity even with a very small
neutral hydrogen fraction. At higher redshifts, the mean density of
the universe is higher, the voids are less empty because structure
formation is still in its infancy, and the neutral fraction is higher
since the ionizing intensity is lower. In addition, if the \ion{H}{2}
region is itself surrounded by still-neutral IGM then the damping
wings of the Ly$\alpha$ line due to this neutral gas can be broad
enough to produce a substantial opacity over the entire wavelength
range corresponding to the \ion{H}{2} region. Indeed, \citet{jordi98}
showed that the neutral IGM can block out all \ion{H}{2} regions up to
a diameter of $\sim 1$ Mpc.

To determine the opacity, we begin with the standard \citep{GP}
opacity for absorption which includes the resonance,
\be
\tau_{\rm GP}={\pi e^2 f_\alpha \lambda_\alpha n_{\HI}(z) \over m_e
cH(z)} = 3.61\times 10^5 \left({\Omega_b h\over 0.0325}
\right)\left({\Omega_m\over 0.35}\right)^{-{1 \over 2}} \left({1+z\over 7}
\right)^{3 \over 2}\ , \label{G-P}
\ee
for neutral gas at the mean density at redshift $z$. Here and
throughout this paper we assume the high-redshift form for Hubble's
constant, \be H(z) \approx H_0 \sqrt{\Omm} \left( 1+z \right)^{3 \over
2}\ , \ee where $H_0$ is Hubble's constant at $z=0$. More generally,
gas at relative density $\Delta = \rho_g /\bar{\rho}_g$ and with a
neutral fraction $x_{\HI}$ produces an optical depth 
\be
\tau_{\rm reson} = \tau_{\rm GP}\, x_{\HI}\, \Delta\ , \ee
where the subscript stands for resonance.

For the density distribution in the IGM within the \ion{H}{2} bubble,
we adopt the general picture (though not the full model) of
\citet{jordi}. This picture is based on the fact that voids can be
quickly and easily ionized while the high recombination rate of dense
clumps keeps them neutral. Thus, a rough description of the state of
the IGM is that most of the gas up to some fiducial overdensity is
ionized, and most of the gas above that overdensity is still
neutral. Although the remaining neutral clumps can contain a
significant gas mass, they take up only a small fraction of the volume
and do not block a significant fraction of the ionizing radiation
emanating from the central source. Thus the neutral clumps can
essentially be ignored, and the total recombination rate is typically
not high enough to limit the size of the \ion{H}{2} bubble. Thus we
describe the IGM within the large \ion{H}{2} bubble as having an
average density $\Delta$ relative to the mean, where it is understood
that the '\ion{H}{2} bubbles' are not fully ionized since they contain
a small volume of dense neutral clumps. Since the clumps are denser
than average, the gas within the bubble which {\it is}\, ionized is
characterized by a $\Delta < 1$. For the model of the density
distribution used in \citet{jordi}, if gas at $z=6$ is ionized up to a
density of 10 times the cosmic mean then the ionized gas has a
$\Delta=0.87$ and a clumping factor (i.e., mean value of density
squared) of $C=1.6$ times that for gas at the cosmic mean density. If
the gas is ionized only up to a density of 2 times the cosmic mean
then the numbers are $\Delta= 0.67$ and $C=0.60$. Because of these
relatively low clumping factors, an ionizing source can easily ionize
the gas up to an overdensity of $\sim 10$ without sacrificing many
photons to balance recombinations. The actual maximum overdensity of
ionized gas depends on the detailed geometry of the overdense regions
and must vary with the distance from the source. Indeed, the model of
\citet{jordi} is most accurate when the ionizing intensity in the IGM
is uniform, and this is true only well after overlap; therefore, in
this paper we instead use a model based on the \ion{H}{2} bubbles of
individual sources, a model which is much more accurate during
overlap. We use the \citet{jordi} picture only in the limited sense of
determining the effective value of $\Delta$. An important detail is
that while the size of the region that a given source can ionize
depends essentially only on the mean $\Delta$ of ionized gas, the
probability of observing a low optical depth in the region is
particularly sensitive to the presence of the lowest-density
voids. For the density distribution of \citet{jordi}, only $0.9\%$ of
the mass, and $4.2\%$ of the volume, is occupied by gas with $\Delta <
0.25$ at $z=6$. The numbers are $7.6\%$ of the mass and $25 \%$ of the
volume, for $\Delta < 0.4$. In our calculations we thus use two
different effective values of $\Delta$; we adopt as standard values
$\langle\Delta\rangle=0.8$ which is used to calculate the size of each
bubble, and $\Delta_{\tau}=0.4$ which is used to calculate the optical
depth of the gas. Thus, e.g., the radius of the \ion{H}{2} bubble
[eq.~(\ref{rmax})] is modified to
\be r_{\rm max}= 82\, 
{\rm kpc} \left(\frac{7}{1+z} \right) \left( \frac{\Ni}{20}\, 
\frac{M} {10^9 M_{\sun}}\, \frac{0.148}{\Omm h^2}\, \frac{0.8}
{\langle\Delta\rangle} \right)^{\frac{1}{3}}\ . \label{rmax2} \ee 

We assume ionization equilibrium for the hydrogen within the
\ion{H}{2} bubble. This is a safe assumption even though the
recombination time for the IGM at $z=6$ is around the Hubble
time. Since equilibrium typically means a neutral fraction of $\sim
10^{-3}$ or less then the time to reach equilibrium is at most
$10^{-3}$ of the recombination time. The recombination rate per volume
is \be \mbox{Rec rate}=\alpha_B\, (\bar{n}_H^0)^2\,
\frac{\Delta^2}{a^6}\ , \ee where $a=1/(1+z)$ and the case B
recombination coefficient for hydrogen at $T=10^4$ K is
$\alpha_B=2.6\times 10^{-13}$ cm$^3$ s$^{-1}$. The ionization rate per
volume is \be \mbox{Ion rate}= x_{\HI}\, \frac{d N_\gamma}{dt}\,
\frac{1}{4 \pi r^2}\, \bar{\sigma}\, \bar{n}_H^0\, \frac{\Delta}{a^3} ,\ee 
where $\bar{\sigma}$ is a frequency-averaged photoionization cross
section, $\sim 2 \times 10^{-18}$ cm$^2$. We set $d N_\gamma/dt =
N_\gamma/t_s$, where $N_\gamma$ is the total number of ionizing
photons produced (which equals the number of baryons in the halo times
$\Ni$) and $t_s$ is the source lifetime (see \S \ref{sec:SourceModel}). 
Then the neutral fraction at radius $r$ in the
\ion{H}{2} bubble is
\be x_{\HI} = 3.15 \times 10^{-3} \left( 
\frac{t_s}{1.3 \times 10^8\mbox{ yr}} \right) \left( \frac{r_{\rm max}}
{82\,{\rm kpc}} \right)^{-1} \left( \frac{\Delta_{\tau}/ \langle\Delta
\rangle}{0.5} \right) \left( \frac{r}{r_{\rm max}} \right)^2\ ,
\ee
and the optical depth at Ly$\alpha$ caused by this gas is therefore
\be
\tau_{\rm reson} = 
455 \left({1+z\over 7} \right)^{9 \over 2} 
\left({\Omega_m h^2\over 0.148} \right)^{1 \over 2} \left( 
{\Omega_b h^2\over 0.0211}\ \frac{t_s}{1.3 \times 10^8\mbox{ yr}}
\right) \left( \frac{\Ni}{20} \frac{M} {10^9 M_{\sun}}\right)^{-1} 
\left( \frac{\Delta_{\tau}}{0.4}\ \frac{r} {82\,{\rm kpc}}
\right)^2\ . \ee
Note that the optical depth at a given radius depends on
$\Delta_{\tau}$ (by definition of $\Delta_{\tau}$), but it does not
depend on $r_{\rm max}$ (and thus on $\langle \Delta \rangle$),
assuming that $r < r_{\rm max}$.

Our choice of duty cycle $\zeta$ affects the statistics of absorption
not only through the lifetime $t_s \propto \zeta$; if, e.g.,
$\zeta=0.25$, then for each active source with a highly ionized
\ion{H}{2} bubble there are three 'dead' bubbles, regions ionized by
sources that have already turned off before $z=6$. These bubbles have
typically been dead for $\sim 2 \times 10^8$ yr, while the
recombination time (if $\langle\Delta\rangle=0.8$) is $\sim 2.5 \times
10^9$ yr. Thus, if the bubbles were initially highly ionized, at $z=6$
they have a neutral hydrogen fraction below $10\%$. This fraction is
high enough that the internal optical depth in these bubbles satisfies
$\tau_{\rm reson} \gg 1$, but the fraction is low enough that this gas
does not contribute significant absorption through the Ly$\alpha$
damping wings. Thus, we treat the dead bubbles as neutral for the
purpose of internal optical depth, but as fully ionized for the
purpose of damping wings and in the calculation of the total filling
factor of ionized gas in the universe.

\subsection{Specific Predictions}

\label{sec:FinModel}

In order to analyze the importance of the observations, we look for a
specific property of the observed spectra that can be computed from
our model. Although the mean optical depth declines with redshift, and
this quantity is straightforward to determine observationally, we
prefer to focus on the essential feature which defines Gunn-Peterson
absorption. This is not the value of the mean optical depth, which is
already rather high even at a somewhat lower redshift; the key feature
is instead the presence of a long stretch of absorption which is
continuous, i.e., which does not allow through any significant flux at
all, even when observed with a very high spectral resolution. This
feature also captures the physical difference between the IGM before
and after reionization. At low redshift, observed spectra are
characterized by overall transmission except for a dense forest of
relatively narrow absorption lines corresponding to the cosmic web of
sheets and filaments. The Ly$\alpha$ forest becomes denser with
redshift, but as long as the low-density gas in the IGM is very highly
ionized, a high-resolution spectrum should always show some
transmission spikes, even when these spikes are rare and the mean
optical depth is high. At redshifts approaching reionization, on the
other hand, the IGM itself has a significant neutral fraction, and it
produces continuous strong absorption over large redshift
intervals. Before full reionization, parts of the IGM are neutral and
they produce even longer intervals of total absorption.

From the previous subsections, our main parameters and their default
values are the redshift $z=6$, the minimum halo circular velocity
$V_c=16.5$ km/s of halos with galaxies, the relative density
$\Delta_{\tau}=0.4$ of ionized gas within the bubbles (where also
$\langle\Delta\rangle=2 \Delta_{\tau}$), the overall source efficiency
$\Ni=20$, and the duty cycle $\zeta=0.25$. For given parameter values,
we calculate the total filling factor $Q_{\rm II}$ of \ion{H}{2}
bubbles, which is simply the integral of halo number density times the
maximum ionized volume $4 \pi r_{\rm max}^3/3$ due to each halo. We
emphasize that the volume taken up by \ion{H}{2} bubbles includes the
small volume fraction of neutral clumps which remain within each
bubble. The end of the overlap stage of reionization corresponds to
$Q_{\rm II}=1$, because at this point all the low-density gas is
ionized. We allow $Q_{\rm II}$ to be greater than unity in the
post-overlap stage, where, e.g., a value of $Q_{\rm II}=5$ means that
the total number density of ionizing photons produced by all sources
is five times greater than the mean number density of hydrogen atoms
in the IGM (where the dense neutral clumps are not included). For
direct comparison with observations, we compute also the mean free
path $\lambda_{\rm GP}$ for observing a region with optical depth less
than a given $\tilde{\tau}$. We deal with the mean free path, a
statistical quantity, since the ionizing intensity in the IGM is
extremely inhomogeneous during reionization. At redshift $z$, we have
\be \lambda_{\rm GP} = \left[ \int_{M_{\rm min}}^{\infty} 
\sigma_{\tilde{\tau}} (M)\, \frac{dn}{dM}\, dM \right]^{-1}\ ,\ee where 
$n(M)$ is the proper number density of halos with mass up to $M$, and
$\sigma_{\tilde{\tau}}(M)$ is the cross-section for intersecting some
point with a Ly$\alpha$ optical depth less than $\tilde{\tau}$ along a
line of sight through the \ion{H}{2} bubble produced by a halo of mass
$M$. The mass $M_{\rm min}$ is determined by the halo with the minimum
$V_c$ or, if it is bigger, by the smallest halo whose \ion{H}{2}
bubble is large enough to produce some point with $\tau <
\tilde{\tau}$. In a stretch of length $l$ along a random line of
sight, the average intersected number of \ion{H}{2} bubbles with a
$\tau < \tilde{\tau}$ region is $l / \lambda_{\rm GP}$. Also, the
chance of not intersecting any such bubble in the same stretch is
exp$[-l / \lambda_{\rm GP}]$. We can similarly calculate a mean free
path for encountering any ionized gas, regardless of the optical
depth:
\be \lambda_{\rm II} = \left[ \int_{M_{\rm min}}^{\infty} 
\sigma_{\rm II}(M)\, \frac{dn}{dM}\, dM \right]^{-1}\ , \label{eq:lII}
\ee where in this case $M_{\rm min}$ is always set by the cutoff $V_c$, 
and $\sigma_{\rm II}(M) = \pi r_{\rm max}^2$. It is also easy to show
that a fraction $Q_{\rm II}$ of a random line of sight is covered by
some part of an \ion{H}{2} bubble, so on average a stretch of length
$\lambda_{\rm II}$ intersects one bubble with a typical intersected
path of length $Q_{\rm II} \lambda_{\rm II}$.

In order to obtain $\lambda_{\rm GP}$, we must first calculate the
optical depth of every parcel of gas. To calculate the optical depth
at each point within an \ion{H}{2} bubble at redshift $z$ we add
$\tau_{\rm reson}$ to the optical depth caused by the red damping
wings of neutral IGM at lower $z$ and the blue damping wings of
neutral IGM at higher $z$. We include the blue damping wings since we
are considering here bubbles along the line of sight to a given source
(e.g., a quasar), and are not considering the region ionized by the
quasar itself (i.e., the proximity effect). The optical depth is given
by an integral along the line of sight of the number density of
neutral hydrogen times the scattering cross-section of the Ly$\alpha$
line. Consider at high redshift a stretch of IGM at the cosmic mean
density which has a neutral fraction $x_{\HI}$ and is located between
redshifts $z_1$ and $z_2$. On light with an observed wavelength
$\lambda =
\lambda_{\alpha} (1+z)$, this gas produces through the damping wing an
optical depth of
\citep{jordi98}
\be \tau_{\rm damp} = 6.43 \times 10^{-9}\, x_{\HI}\, \tau_{\rm GP} 
\left[ {\rm I}\left(\frac{1+z_2} {1+z}\right) - {\rm I}
\left(\frac{1+z_1} {1+z} \right) \right]\ , 
\ee where $\tau_{\rm GP}$ is given by eq.\ (\ref{G-P}) and
\be {\rm I}(x)\equiv {x^{9/2}\over 1-x}+{9\over 7}x^{7/2}+{9\over 5}
x^{5/2}+ 3 x^{3/2}+9 x^{1/2}-{9\over 2} \ln\left| {1+x^{1/2}\over
1-x^{1/2}} \right|\ , \ee and we have assumed $z < z_1 < z_2$ or $z_1
< z_2 < z$ so that the absorption does not involve the central region
of the Ly$\alpha$ line, near resonance.

The optical depth produced by an IGM with a given filling factor
$Q_{\rm II}$ depends on the distribution of neutral and ionized
regions within the IGM. When $Q_{\rm II} \ge 1$ we assume that there
is no neutral IGM and that $\tau_{\rm damp}=0$. In this case, not only
is there no neutral IGM, but also most points in the IGM feel the
radiation from more than one source. As a simple way of accounting
approximately for the high number density of sources, for each source
we consider its bubble only out to radius $r_{\rm max}/\sqrt{Q_{\rm
II}}$, since the more distant gas is likely to be closer to some other
source, given that $Q_{\rm II} > 1$. Note that we are still assuming
that the radiation intensity at each point is dominated by the nearest
source; we discuss this issue further in \S \ref{sec:Disc}. 

When $Q_{\rm II} < 1$, $\tau_{\rm damp}$ depends on the exact
distribution of neutral and ionized gas on both sides of a given
\ion{H}{2} bubble.  It is not possible to analytically describe the
full complexity of this distribution, so we use the statistical
properties of this distribution in order to get a handle on the
expected range of values of $\tau_{\rm damp}$. Specifically, we know
the volume-averaged neutral fraction $x_{\HI} = 1-Q_{\rm II}$, and we
also know that on average a stretch of length $\lambda_{\rm II}$
intersects a single bubble with a typical intersected path of length
$Q_{\rm II} \lambda_{\rm II}$ [see equation (\ref{eq:lII})]. We
therefore adopt the following simple model for the IGM on each side of
the \ion{H}{2} bubble, along a given line of sight: We take two
consecutive segments of length $\lambda_{\rm II}$ each, where in each
segment the central length of $Q_{\rm II} \lambda_{\rm II}$ is ionized
and the rest is neutral. After these two segments, we simply take a
uniform IGM with the neutral fraction set equal to $1-Q_{\rm II}$. The
optical depth is only weakly dependent on the precise redshift at
which this neutral IGM is cut off, so we fix a minimum $z=4$ and a
maximum $z=12$ (i.e., we do not consider sources observed at $z>12$).
By trying other possible arrangements of neutral gas within the IGM we
find a typical systematic uncertainty of around a factor of two in
$\lambda_{\rm GP}$ (when $Q_{\rm II} < 1$); the particular arrangement
of segments that we have chosen is arbitrary, except that it gives
results that lie in the central range of values obtained from various
possible arrangements.

To illustrate some numerical values, since with our standard
parameters $Q_{\rm II}=5.0$ we consider here an example with $z=8$,
$\Ni = 10$ and $V_c=40$ km/s, for which $Q_{\rm II}=0.39$. In this
case, $\lambda_{\rm II}=0.29$ Mpc. Consider, for example, a $10^{11}
M_{\sun}$ halo, which produces an $r_{\rm max}=0.24$ Mpc; if a line of
sight passes at a projected radius of 0.05 Mpc from the center of this
bubble, the total length through the \ion{H}{2} bubble is 0.46 Mpc. At
the center of this line through the bubble, the optical depth is 11.1,
made up of $\tau_{\rm reson}=5.6$ and $\tau_{\rm damp}=2.7$ and 2.8
from the red and blue damping wings, respectively. Note that in
calculating the damping wings we have neglected the possibility that
they could be affected by density fluctuations or peculiar velocities
in the neutral gas. The overall $\tau_{\rm damp}$ results mainly from
the integrated effect of a length of $\sim 0.5$ Mpc of neutral gas. At
$z=6$, the typical $1-\sigma$ density fluctuation on this scale is
0.32, so fluctuations are likely to have only a moderate effect on the
overall statistics, but the effect is significant and should be
studied. The damping wings block out small bubbles and imply that
typically, when $Q_{\rm II} < 1$, transmitted flux is possible only
through bubbles produced by halos of mass at least $\sim
10^{11}$--$10^{12} M_{\sun}$. This permits a rough estimate of
$\lambda_{\rm GP}$ in this case. Indeed, if $\zeta=0.25$ then at $z=6$
the number density of active bubbles produced by halos with $M>10^{11}
M_{\sun}$ is $\sim 0.1$ per Mpc$^3$. Each bubble has an $r_{\rm max}
\sim 0.4$ Mpc, so if each bubble produces a low optical depth out to a
projected radius of $r_{\rm max}/2$, then $\lambda_{\rm GP} \sim 80$
Mpc. This explains the high values of $\lambda_{\rm GP}$ found in the
next section whenever $Q_{\rm II} < 1$.

\section{Results}

\label{sec:Results}

\begin{figure} 
\plotone{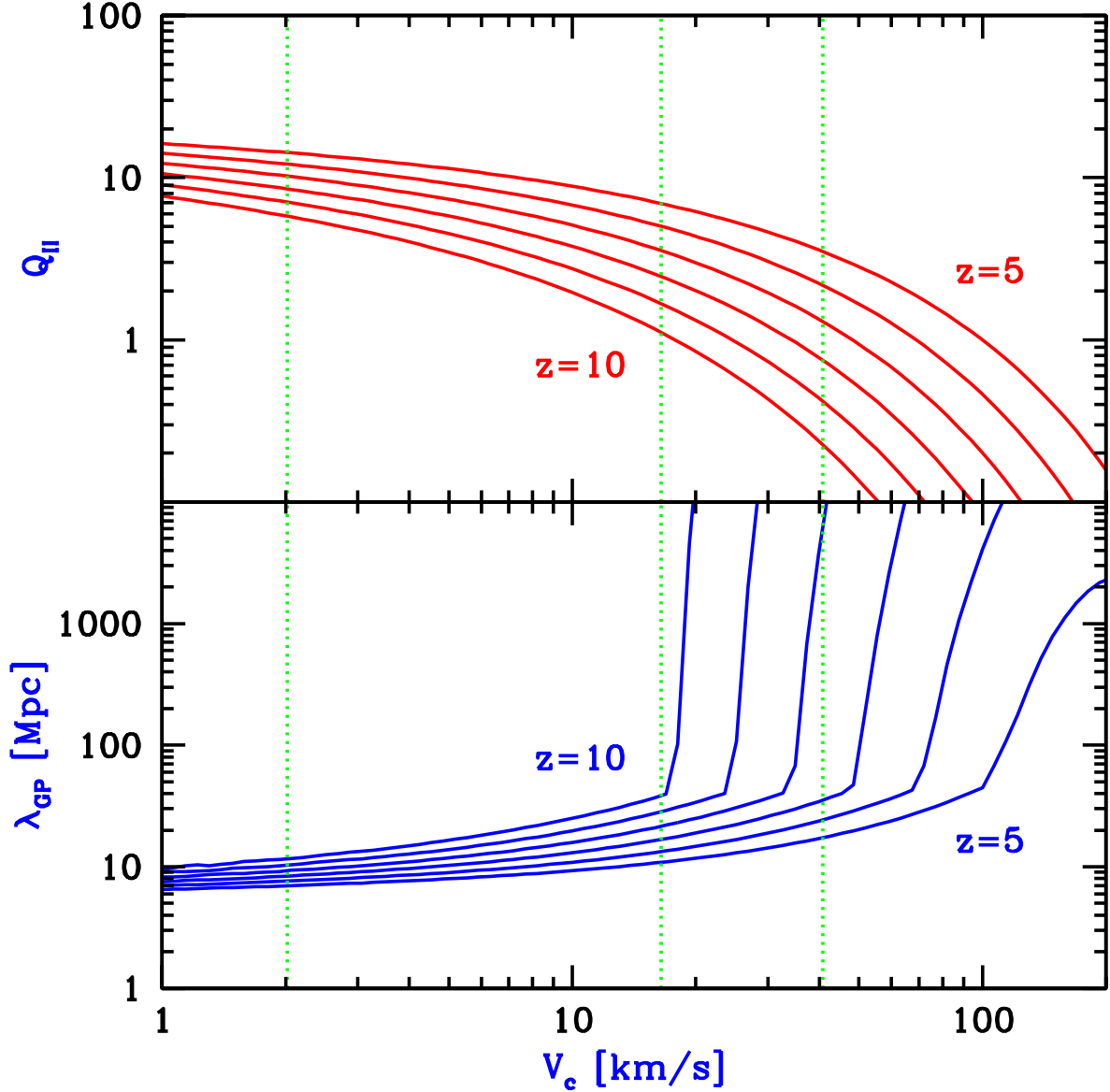} 
\caption{The mean free path $\lambda_{\rm GP}$ and the filling factor
$Q_{\rm II}$ versus the minimum circular velocity $V_c$ for halos
which host galaxies. The bottom panel shows the mean free path for
crossing through a region with a Ly$\alpha$ optical depth less than
$\tilde{\tau}=2.5$. The mean free path in proper Mpc is shown versus
the cutoff, minimum halo circular velocity $V_c$ in km/s. The upper
panel shows the \ion{H}{2} filling factor versus $V_c$. In both
panels, the different curves correspond to redshifts 5, 6, 7, 8, 9,
and 10, as indicated. Vertical dotted lines show the values of $V_c$
that correspond roughly to $H_2$ cooling (2.0 km/s), atomic cooling
(16.5 km/s), and pressure suppression (41 km/s). Other parameter
values are $\Delta_{\tau}=0.4$, $\langle\Delta\rangle=0.8$, $N_{\rm
ion}=20$, and $\zeta=0.25$.}
\label{fig:mfpQVc}
\end{figure}

To compare our model with the observations, we first note the sizes of
the widest dark regions found in the quasar spectra. \citet{z6.3}
found a dark region at $z=5.87$--6.15, or 16.8 Mpc (considering just
Ly$\alpha$, not Ly$\beta$), and \citet{z5.8} found a dark region at
$z=5.25$--5.31, or 4.7 Mpc. In both cases, the highest transmission
spikes within these regions correspond to $\tau \sim 2$, but
measurement noise may well explain these spikes. More quantitatively,
within the dark region \citet{z5.8} estimate a mean flux level of $1.6
\times 10^{-3}$ relative to the quasar continuum, and a standard
deviation of $\sim 0.01$ per resolution element (which is $\sim 10$
pixels). Since there are $\sim 50$ resolution elements within the dark
interval, both of these numbers imply that the upper limit on the flux
within a single resolution element corresponds to $\tau \sim 2.5$. As
our default value we adopt this limit of $\tilde{\tau}=2.5$, i.e., we
assume that even a small stretch with $\tau < 2.5$ can be excluded
within the dark regions in both of the observed spectra. Figure
\ref{fig:mfpQVc} shows the predicted values of $\lambda_{\rm GP}$ and
$Q_{\rm II}$ as a function of the cutoff halo circular velocity $V_c$,
for several redshifts between $z=5$ and $z=10$. For our standard
parameters ($V_c=16.5$ km/s at $z=6$), we find $\lambda_{\rm GP}=13.3$
Mpc and $Q_{\rm II}=5.0$.  At a given redshift, $Q_{\rm II}$ declines
more steeply with $V_c$ at high values of $V_c$, because the halo mass
function declines steeply at high mass values which correspond to very
rare halos. At low $V_c$, $\lambda_{\rm GP}$ depends only weakly on
the cutoff since even when $Q_{\rm II}>1$ the total cross-section of
low-$\tau$ regions in
\ion{H}{2} bubbles is dominated by bubbles due to relatively massive
halos. When the minimum $V_c$ is increased, as $Q_{\rm II}$ drops
below 1 there is a steep rise in $\lambda_{\rm GP}$ due to the effects
of the damping wings of the neutral IGM.

\begin{figure} 
\plotone{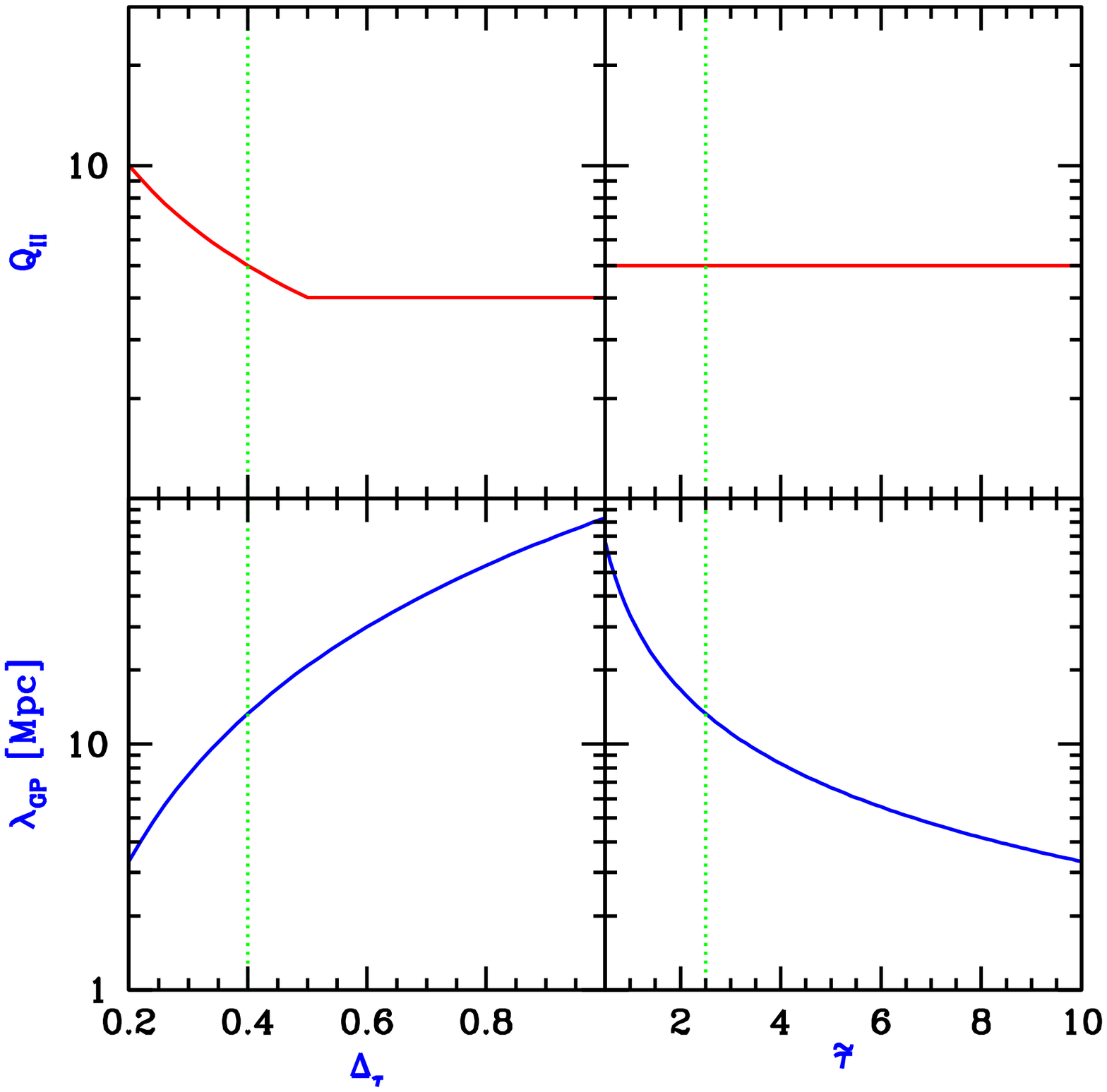} 
\caption{The mean free path $\lambda_{\rm GP}$ and the filling factor
$Q_{\rm II}$. Bottom panels show the mean free path in proper Mpc for
crossing through a region with $\tau < \tilde{\tau}$. Upper panels
show the \ion{H}{2} filling factor. In the left panels, the gas
density $\Delta_{\tau}$ is varied, where $\Delta=\rho/\bar{\rho}$. In
every case we set $\langle\Delta\rangle=2 \Delta_{\tau}$, except that
$\langle\Delta\rangle$ is not increased past unity. The vertical
dotted line shows our standard value of $\Delta_{\tau}=0.4$. In the
right panels, the maximum optical depth $\tilde{\tau}$ is varied. The
vertical dotted line shows our standard value of $\tilde{\tau}=2.5$.
Other parameter values are $z=6$, $V_c=16.5$ km/s, $N_{\rm ion}=20$,
and $\zeta=0.25$.}
\label{fig:mfpQTau}
\end{figure}

Figure \ref{fig:mfpQTau} shows the dependence of $\lambda_{\rm GP}$
and $Q_{\rm II}$ on $\Delta_{\tau}$ and on $\tilde{\tau}$. In every
case we set $\langle\Delta\rangle=2 \Delta_{\tau}$, except that
$\langle\Delta\rangle$ is not increased past unity. In general, a
lower $\langle\Delta\rangle$ makes it easier to reach the end of
overlap, since a greater mass fraction of the gas remains in neutral
clumps which do not need to be ionized in order to reach $Q_{\rm
II}=1$. A lower $\Delta_{\tau}$ lowers $\tau_{\rm reson}$ at a given
radius in a bubble in proportion to $\Delta_{\tau}^2$, where one
factor of $\Delta_{\tau}$ is due to the reduced neutral fraction. The
plot of $\lambda_{\rm GP}$ as a function of $\tilde{\tau}$ indicates
that a more detailed comparison should be possible in the future
between theory and observation. Indeed, the full probability
distribution of $\tau$ as a function of redshift, and the correlation
function of $\tau$ along the line of sight, should be compared between
detailed observations and detailed numerical simulations with accurate
radiative transfer.

\begin{figure} 
\plotone{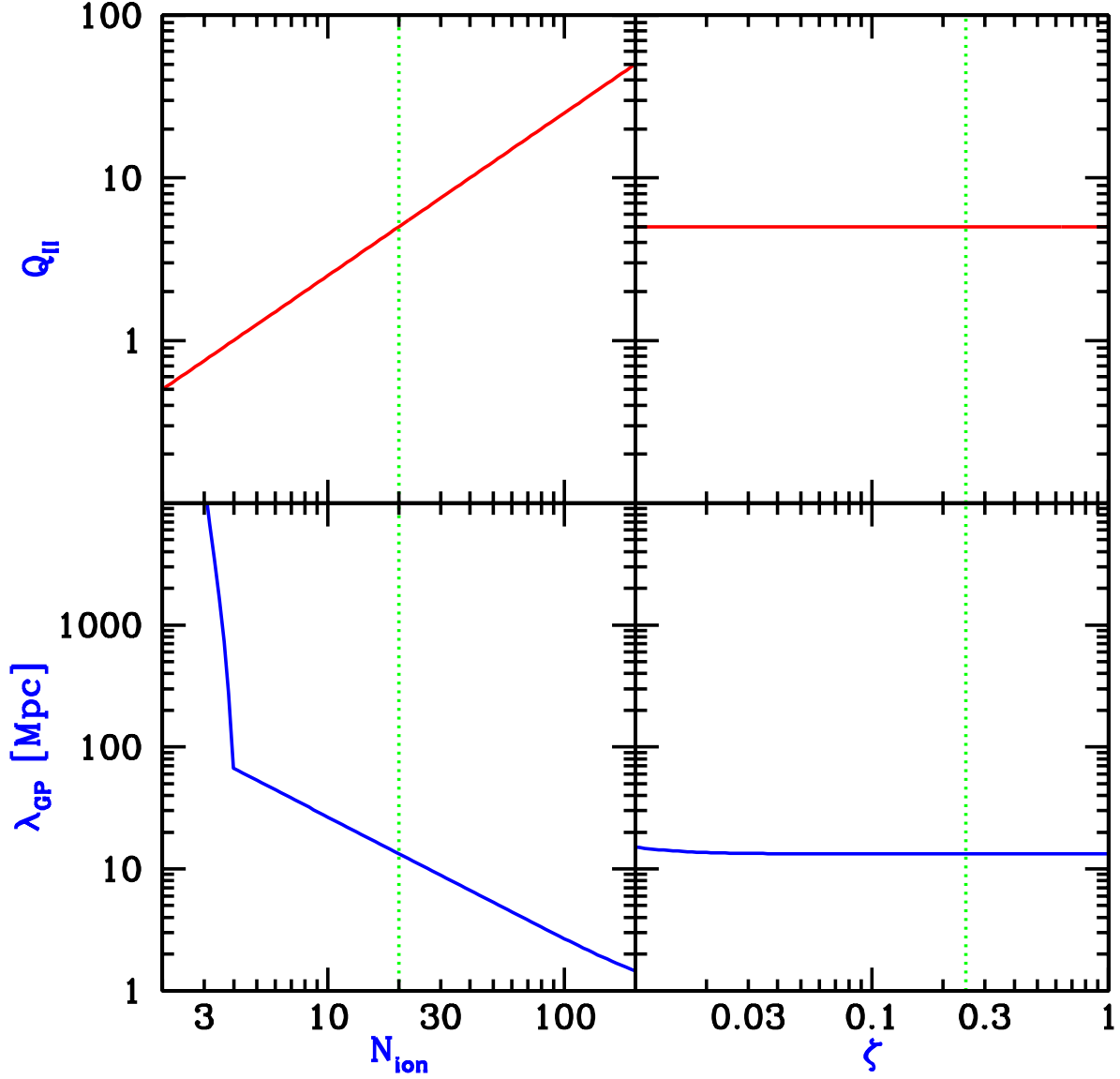} 
\caption{The mean free path $\lambda_{\rm GP}$ and the filling factor
$Q_{\rm II}$. Bottom panels show the mean free path in proper Mpc for
crossing through a region with a Ly$\alpha$ optical depth less than
$\tilde{\tau}=2.5$. Upper panels show the \ion{H}{2} filling
factor. In the left panels, the source efficiency $N_{\rm ion}$ is
varied. The vertical dotted line shows our standard value of $N_{\rm
ion}=20$. In the right panels, the source duty cycle $\zeta$ is
varied. The vertical dotted line shows our standard value of
$\zeta=0.25$. Other parameter values are $z=6$, $V_c=16.5$ km/s,
$\Delta_{\tau}=0.4$, and $\langle\Delta\rangle=0.8$.}
\label{fig:mfpQduty}
\end{figure}

Figure \ref{fig:mfpQduty} shows the dependence of $\lambda_{\rm GP}$
and $Q_{\rm II}$ on the source parameters $\Ni$ and $\zeta$.
Increasing $\Ni$ causes a proportional increase in $Q_{\rm II}$, since
a greater number of ionizing photons is produced. At the same time,
$\lambda_{\rm GP}$ decreases since the ionizing intensity increases at
a given distance from a source. As before, $\lambda_{\rm GP}$
increases sharply if $\Ni$ is lowered to the point where $Q_{\rm
II}<1$. The duty cycle does not affect $Q_{\rm II}$ since we assume
that the bubbles of dead sources remain ionized. Each active bubble is
more highly ionized when $\zeta$ is small, but only bubbles with
active sources can have a low optical depth, so a low $\zeta$ leads to
a slight increase in $\lambda_{\rm GP}$. The dependence on $\zeta$ is
weak also when parameters are chosen so that $Q_{\rm II} < 1$. Given
this weak dependence, $Q_{\rm II}$ and $\lambda_{\rm GP}$ are also
expected to be insensitive to clustering of sources. Consider a simple
toy model for clustering in groups of $N>1$ halos, where we make each
source $N$ times brighter but assume $N$ times fewer sources. As
before, $Q_{\rm II}$ remains unchanged, but each bubble becomes bigger
and more highly ionized at a given radius. Indeed, the radius at which
we find a given $\tau_{\rm reson}$ goes up as $\sqrt{N}$, so the
cross-section goes as $r^2 \propto N$, which cancels out the $1/N$
decrease in the density of sources. Thus, when $Q_{\rm II} > 1$
clustering has almost no effect on $\lambda_{\rm GP}$, but it does
decrease $\lambda_{\rm GP}$ somewhat when $Q_{\rm II} < 1$, since
damping wings have a smaller effect on larger bubbles. Note that even
parameters that do not change $\lambda_{\rm GP}$ significantly {\it
can}\, be probed with detailed observations since they do change
ionized gas fractions within \ion{H}{2} bubbles as well as the
distribution of bubble sizes.

\begin{figure} 
\plotone{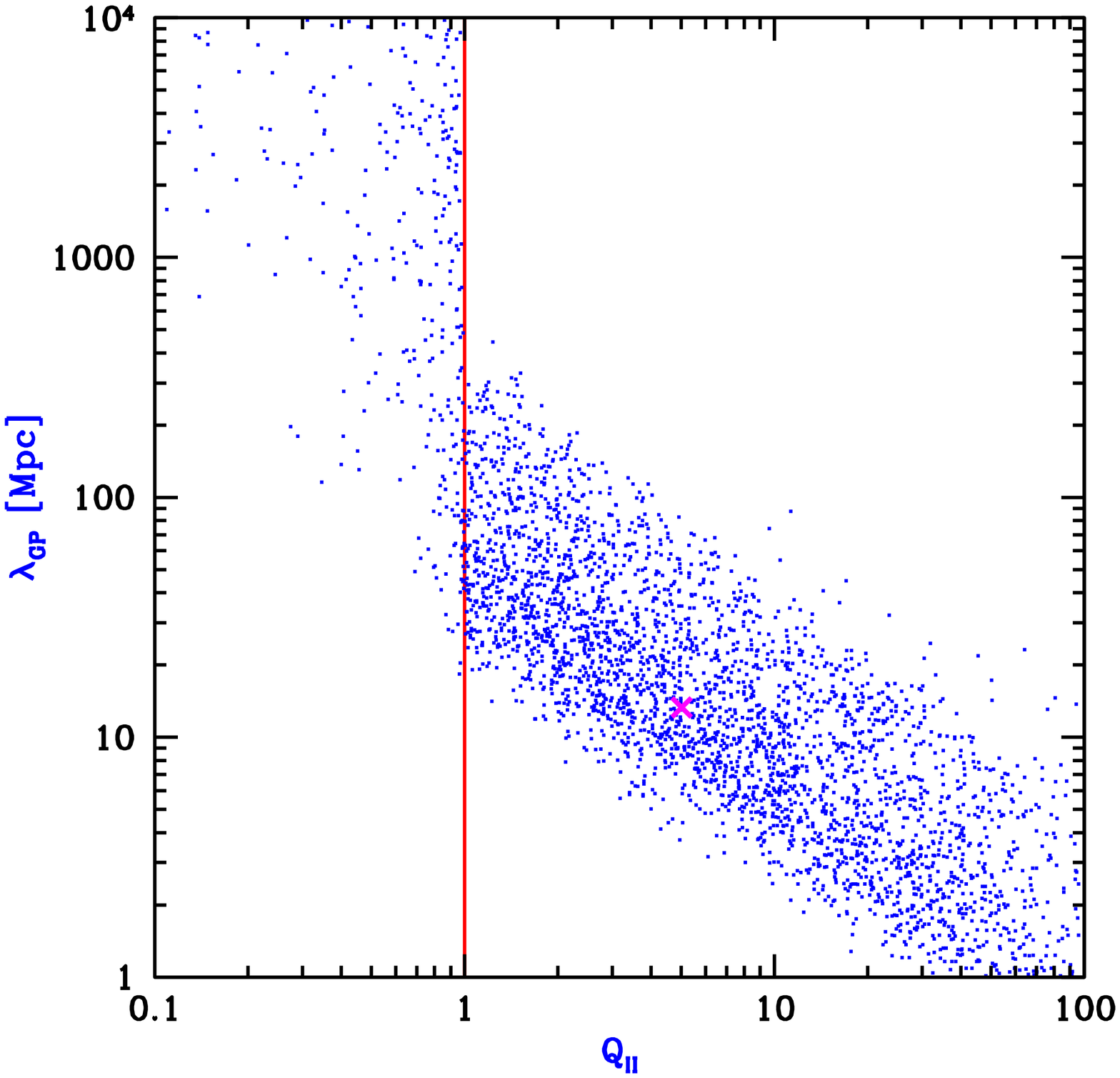} 
\caption{The mean free path $\lambda_{\rm GP}$ versus the filling factor
$Q_{\rm II}$. The vertical axis is the mean free path in proper Mpc
for crossing through a region with a Ly$\alpha$ optical depth less
than $\tilde{\tau}=2.5$. The horizontal axis shows the \ion{H}{2}
filling factor. The scattered points show the results from 5000 sets
of parameters selected randomly from the parameter ranges shown in
figures \ref{fig:mfpQVc}, \ref{fig:mfpQTau}, and \ref{fig:mfpQduty}
(except that $\tilde{\tau} = 2.5$ is held fixed). The large $\times$
marks the point corresponding to our standard parameter values. The
solid vertical line is drawn at $Q_{\rm II}=1$.}
\label{fig:mfpQ}
\end{figure}

The discussion so far shows that after considering a wide range of
parameter values we expect the universe to be well past the end of
overlap by redshift six. However, present observations place only weak
direct constraints on the parameters controlling the abundance and
efficiency of sources at such high redshifts. Thus, our goal is to use
directly the observations of wide regions of Gunn-Peterson absorption
to infer the state of reionization at $z=6$, regardless of the precise
parameter values. To this end, Figure \ref{fig:mfpQ} shows a scatter
plot of points derived from the entire parameter space of reasonable
values. Shown are 5000 points, where each corresponds to a set of
parameters selected randomly from the ranges $z=5$--10, $V_c=1$--200
km/s, $\Delta_{\tau}=0.2$--1, $\Ni$=2--200, and $\zeta=0.01$--1. We
hold $\tilde{\tau} = 2.5$ fixed, and choose parameters from uniform
distributions in the log (except for $z$). In the figure we also mark
with an $\times$ the location corresponding to our standard parameter
values, and we draw a straight line at $Q_{\rm II}=1$. The figure
clearly illustrates that over a broad range of possible parameter
values, there is a very strong correlation between $Q_{\rm II}$ and
$\lambda_{\rm GP}$. Indeed, for most parameter values we find that
$\lambda_{\rm GP} < 100$ Mpc if and only if $Q_{\rm II} > 1$.  Thus,
when $Q_{\rm II} < 1$ the incidence of transparent regions should
become very low, since for example $z=6$--7 corresponds to a
line-of-sight length of only 51 Mpc. Thus, e.g., if $\lambda_{\rm
GP}=1000$ Mpc then the chance of encountering any $\tau < 2.5$ spot
along a given line of sight is about 1 in 20. Note also that while the
$y$-axis in the figure is shown only up to $\lambda_{\rm GP} = 10^4$
Mpc, most of the points at $Q_{\rm II} < 1$ actually lie at even
higher values of $\lambda_{\rm GP}$.

The appearance of Figure \ref{fig:mfpQ} is determined by the various
input parameters whose individual effects were shown in the previous
figures. If $V_c$ is varied from small to large values, with all other
parameters fixed, the result is a single curve which goes from the top
left to the bottom right corner in the figure. Such curves drawn at
various redshifts all lie almost on top of each other at $Q_{\rm II} >
1$, but at $Q_{\rm II} < 1$ the redshift dependence explains almost
all of the scatter in the figure; in fact, if we used only $z>6$
instead of $z>5$, there would be significantly less scatter at $Q_{\rm
II} < 1$, with the points sticking closer to the vertical line $Q_{\rm
II} = 1$. In order to be conservative we also varied $\Delta_{\tau}$
over a wide range, and this parameter is responsible for most of the
scatter at $Q_{\rm II} > 1$. As for the other parameters, variations
in the source efficiency $N_{\rm ion}$ can be responsible for only a
minor fraction of the apparent scatter; finally, varying $\zeta$ has
little effect on the figure.

By any reasonable definition of pre-overlap, bubbles of transmitted
flux should be extremely rare during pre-overlap. For a given filling
factor $Q_{\rm II} < 1$, the probability that a given point is inside
at least one bubble is $Q_{\rm II}$, if the point is chosen uniformly
over volume. If we make the approximation that different bubbles are
independently placed then the probability of a given point being in
exactly one bubble is $Q_{\rm II} \exp[-Q_{\rm II}]$. Thus, the
probability that a given point is in two or more overlapping bubbles
is
\be {\rm P}_{\rm over} = Q_{\rm II} \left[ 1-e^{-Q_{\rm II}} \right]\ . \ee
Pre-overlap can be defined as the time when ${\rm P}_{\rm over} \ll
1$. For example, ${\rm P}_{\rm over} = 0.1$ when $Q_{\rm II}=0.34$ and
${\rm P}_{\rm over} = 0.01$ when $Q_{\rm II}=0.10$.

\section{Discussion}

\label{sec:Disc}

In the previous section we showed that over a broad range of possible
parameters that characterize the IGM and the population of ionizing
sources, there is a strong correlation between the reionization state
of the IGM as measured by $Q_{\rm II}$ and the typical observed length
of Gunn-Peterson absorption $\lambda_{\rm GP}$. We now consider the
recent spectral observations in light of these results. Both the 16.8
Mpc dark region at $z=6$ \citep{z6.3} and the 4.7 Mpc dark region at
$z=5.3$ \citep{z5.8} most likely correspond to $Q_{\rm II} > 1$. If
indeed true, this means that these observations correspond to
post-overlap just like observations at lower redshift, but their
novelty comes from the approach toward overlap which implies that the
voids, while being mostly ionized, still contain a high enough neutral
fraction to produce wide regions of high optical depth in spectra.
The $z=5.3$ observation cannot correspond to $Q_{\rm II} < 1$, since
there are regions of high transmitted flux at still higher redshifts
in the same spectrum, while $Q_{\rm II}$ is expected to increase with
time (see Figure \ref{fig:mfpQVc}), and $Q_{\rm II} < 1$ should
correspond to a spectrum that is almost completely dark except for the
proximity effect of the quasar (see Figure \ref{fig:mfpQ}). Note also
that we have selected the 4.7 Mpc region at $z=5.3$ as the widest dark
region over the range $z=5.2$--5.6, which measures 30 Mpc. This
selection effect means that the length of 4.7 Mpc must be a few times
larger than the effective $\lambda_{\rm GP}$ over this redshift range.
In contrast, the $z=6$ observation may in fact imply the discovery of
the IGM during overlap, since the dark region is observed right up to
the region affected directly by the quasar emission.  Clearly,
however, a dark region of $\sim 50$ Mpc must be observed before we can
conclude that we are likely observing the effects of damping wings and
of the $Q_{\rm II} < 1$ era. Alternatively, if 15 Mpc dark regions are
observed at $z=6$ along many lines of sight, with the dark regions
always extending right up to the zone of the proximity effect of the
quasar, this also will constitute strong evidence that the end of
overlap has been observed.

We now consider an additional argument which deals with an aspect of
reionization that has not been included thus far. We examine the
effect of the increasing intensity due to multiple sources reaching
every point within overlapping bubbles. This also has implications for
the amount of time after the end of overlap during which wide regions
of Gunn-Peterson absorption can still be observed. We embark on this
discussion in order to understand an additional implication of the
recent observations, which is a result of the significant differences
between the spectra along the two lines of sight. We argue that the
observations are inconsistent with a rapid, homogeneous overlap era,
and that this presents an important challenge to theoretical models.

The end of overlap ($Q_{\rm II}=1$) marks the first time when all gas
in voids can see at least one ionizing source. Thus, as long as
$Q_{\rm II} < 1$, some voids are still completely neutral and they
block the radiation from some ionizing sources. The last stages of
overlap should occur within a small fraction of a Hubble time at the
overlap redshift, since each source is required to ionize only the
region out to its nearest neighbors, which are likely to be fairly
nearby. In addition, the photons which gradually succeed in ionizing
the dense surroundings of each galaxy during pre-overlap then find it
very easy to ionize the low-density voids during overlap. Each time
two or more bubbles are joined, the gas within the combined volume is
exposed to the ionizing radiation from more then one
source. Therefore, the ionizing intensity inside \ion{H}{2} regions
begins to rise, allowing those regions to expand into high-density gas
which had previously recombined fast enough to remain neutral when the
ionizing intensity had been low. Since each bubble coalescence
accelerates the process of reionization, the overlap phase has the
character of a phase transition and should occur rapidly.

At the end of overlap, the final shadows cast by neutral gas in voids
disappear, and every point in the IGM can begin to see all ionizing
sources in the universe, except that two barriers must still be
overcome. One barrier is simply due to the light travel time. At
redshift $z$, the ionizing intensity at a given point is determined by
taking the number density of sources times the flux of each source,
and integrating over a spherical volume element out to a radius
$r$. The flux of each source is given by its intensity divided by
$r^2$. The intensity of each source is proportional to the halo mass,
times an efficiency factor, divided by the source lifetime. If we use
for the purposes of this estimate the approximate assumption that all
sources have the same lifetime, then the result depends only on an
integral of halo mass times number density, i.e., on the total
collapse fraction $F_{\rm col}$, which is the mass fraction contained
in halos which host galaxies. Note that we can set $\zeta=1$ since a
smaller duty cycle increases the intensity of each source but
decreases the number density of sources by the same factor. For gas of
relative density $\Delta_{\tau}$, the optical depth which results from
integrating sources out to radius $r$ is
\be \tau_{\rm reson} = 9.78 \left({1+z\over 7} \right)^{3 \over 2} 
\left({\Omega_m h^2\over 0.148} \right)^{-{1 \over 2}} \left( 
{\Omega_b h^2\over 0.0211}\ \frac{t_s/\zeta}{5 \times 10^8\mbox{ yr}}
\right) \left( \frac{\Ni}{20} \frac{F_{\rm col}} {0.15}\right)^{-1} 
\left( \frac{\Delta_{\tau}}{0.4}\right)^2 \left( \frac{r}{\rm Mpc} 
\right)^{-1}\ . \ee

The result that the optical depth approaches 0 when $r$ is large
arises from the proportionality of the total intensity to the radius
$r$. This divergence of intensity with distance is well known from
Olber's paradox, which applies here since we are considering a small
region and ignoring redshift effects. This simple estimate implies
that if the light travel time were the only barrier, gas in voids
would reach $\tau < 2.5$ in 13 Myr, and $\tau < 1$ in 32 Myr. These
numbers can be increased by making sources less efficient (i.e.,
decreasing $\Ni$) or by making high-mass halos dominate over low-mass
ones. As mentioned in \S \ref{sec:SourceModel}, one natural process
which increases $V_c$ is the suppression of gas infall into low-mass
halos. This suppression may already be important at the end of
overlap, and if it raises $V_c$ from 16.5 km/s to 41 km/s then $F_{\rm
col}$ at $z=6$ decreases from 0.15 to 0.066, increasing the light
travel time needed to reach $\tau < 2.5$ by a factor of 2.3. If future
observations discover that star formation at high redshift occurs
efficiently only in galaxies with $V_c=100$ km/s or greater, then the
light travel time will be increased by an additional factor of 4.7
beyond the effects of pressure suppression. In addition, although we
have used the fiducial value $\Delta_{\tau}=0.4$ in the estimate of
the optical depth, in this context we are particularly sensitive to
voids which have a relatively low underdensity and therefore still
produce wide regions of Gunn-Peterson flux. If we find a void with
$\Delta_{\tau}=0.8$, then in this region the light travel time is
increased by a factor of 4.

The second barrier to increasing the ionizing intensity comes from the
shadow due to the small volume of gas which remains in neutral clumps
even during post-overlap. As in the discussion at the beginning of \S
\ref{sec:IGMModel}, we adopt here the model of \citet{jordi} for the
density distribution in the IGM and also for the resulting mean free
path of ionizing photons. In this model, if gas at $z=6$ is ionized up
to a density of 10 times the cosmic mean then the ionized gas has a
clumping factor of $C=1.6$ times that for gas at the cosmic mean
density, and the mean free path is 2.9 Mpc. If the gas is instead
ionized up to 100 times the cosmic mean density then $C=4.5$ and the
mean free path is 33 Mpc. Thus, the remaining shadow of neutral
regions may temporarily prevent the intensity from increasing rapidly,
at least until most of the gas in filaments with moderate
overdensities is ionized. A different way to get a similar result is
to extrapolate to high redshift the absorption due to the Ly$\alpha$
forest; \citet{z6.3} find that this implies an overall average optical
depth of $\sim 2$--3 at $z=5.3$, and this fairly high value implies a
shadow that must keep the ionizing intensity from increasing rapidly.

An important role in delaying the end of the Gunn-Peterson trough may
be played by the population of photo-evaporating halos. If molecular
hydrogen is dissociated, as discussed above, then galaxies around the
time of overlap form only in halos above $\sim 10^8 M_{\sun}$, where
atomic cooling is efficient. However, most of the gas which lies in
halos is inside smaller halos (down to $\sim 10^5 M_{\sun}$), where
the lack of cooling prevents the formation of galactic disks and stars
or mini-quasars. \citet{me99} showed that photoionization heating by
the cosmic UV background \citep[or shock heating:][]{cen01} could then
evaporate much of this gas back into the IGM, with the process
beginning during overlap. They showed that this process affects a
broad range of halo masses, with only a small gas fraction evaporating
out of $10^8 M_{\sun}$ halos, but with halos below $\sim 10^6
M_{\sun}$ losing their entire gas content because of their shallow
gravitational potential wells. For overlap at $z \sim 6$, around
$20\%$ of the gas in the universe undergoes this photo-evaporation
process, and this gas represents most of the gas which is already in
halos just before overlap. Gas in the smallest halos evaporates very
quickly into the IGM, but the larger halos which retain some of their
gas expand only gradually, and may cast a shadow for some time,
preventing ionizing photons from reaching large distances. These
surviving halos may also contribute to the high column density end of
the Ly$\alpha$ forest at $z \sim 3$
\citep{bss88,am98}.

The process of reionization may be very inhomogeneous, with overlap
occurring at substantially different times in different places. Such
inhomogeneity can be enhanced due to variations in the IGM density,
clustering of sources, and spatial or temporal variation in source
parameters such as the efficiency $\Ni$. Still, a large inhomogeneity
is unlikely if the relatively common low-mass halos dominate. Taking
the cooling mass as the lower mass limit at $z=6$, for instance, there
are 3000 halos per Mpc$^3$. The number density is 130 per Mpc$^3$ for
$V_c=41$ km/s, and 3 per Mpc$^3$ for $V_c=100$ km/s. Thus, if the
efficiency $\Ni$ is high only in high-mass halos, then strong
Gunn-Peterson absorption can last down to lower redshift, because
there are fewer stars in galaxies (i.e., $F_{\rm col}$ is smaller),
because reionization is more inhomogeneous, and because the bubble
around each individual source must grow very large and thus the source
must overcome a strong shadow due to neutral clumps within the bubble.
Whatever the explanation, the two recent observations together imply
that the increase in the ionizing intensity during post-overlap must
have been gradual, or that overlap must have occurred
inhomogeneously. This follows from the fact that overlap in one line
of sight occurs at $z > 5.9$, while a relatively large region of
strong absorption is found along a second line of sight at $z=5.3$,
and the age difference between these redshifts is large (140 Myr).
This significant cosmic variance, which has been observed even with
only two lines of sight, further justifies the statistical approach
that we adopted in \S \ref{sec:FinModel}, where we opted to study
$\lambda_{\rm GP}$ rather than the mean optical depth.

Note that the the ionizing intensity starts to increase in overlapping
bubbles even when $Q<1$, but this effect should be significant only in
the short period when $Q$ approaches unity. Furthermore, when $Q$ is
significantly less than unity any transmitted flux must be very rare
since the damping wings block out flux from most bubbles very
effectively, almost regardless of the source parameters.

\section{Conclusions}

\label{sec:Conc}

We have constructed a model of reionization in order to interpret
recent observations of strong absorption in quasar spectra. By taking the
\ion{H}{2} bubble produced by each ionizing source, and considering
the statistics of the ensemble of these bubbles, we have derived for
given source parameters the filling factor $Q_{\rm II}$ of \ion{H}{2}
in the IGM, and the mean free path $\lambda_{\rm GP}$ for observing a
region with optical depth less than a given $\tilde{\tau}$. In
particular, we expect the universe to be well past the end of overlap
by redshift six, with a $\lambda_{\rm GP} \sim 10$ Mpc. However,
regardless of the source parameters, we find a strong correlation
between the reionization state of the IGM as measured by $Q_{\rm II}$
and the typical observed length of Gunn-Peterson absorption
$\lambda_{\rm GP}$. We conclude that the post-overlap era is
consistent with the observed dark regions of 16.8 Mpc at $z=6$
\citep{z6.3} and 4.7 Mpc at $z=5.3$ \citep{z5.8}.  The $z=5.3$
observation cannot correspond to $Q_{\rm II} < 1$, since the universe
before the end of overlap should correspond to a spectrum that is
almost completely dark except for the proximity effect of the
quasar. This strong absorption results from the damping wings of
neutral IGM, which block out the flux from all except very large, and
rare, \ion{H}{2} bubbles. The $z=6$ observation is consistent with the
state of the IGM before the end of overlap, since the dark region
extends right up to the highest available redshift. If this
interpretation is the correct one then much wider dark regions are
expected and should be revealed by spectra of even higher-redshift
sources. Alternatively, the discovery of the reionization era can be
established even with quasars at $z \ga 6$ if multiple lines of sight
all show Gunn-Peterson absorption.

Combining the two recent observations constrains the evolution of the
ionizing intensity during and after overlap. The observations imply
that along one line of sight, overlap occurs at $z > 5.9$, while a
relatively large region of strong absorption is found along a second
line of sight at $z=5.3$. The age difference between these redshifts
is 140 Myr, which is much longer than our estimate of the light travel
time needed to make the voids transparent due to the combined
intensity of multiple ionizing sources. Thus, the ionizing intensity
during post-overlap must have increased gradually, with a delay caused
by the shadows due to the remaining neutral clumps. The neutral gas in
halos which photo-evaporated during and after overlap may also have
cast a significant shadow. If reionization was dominated by relatively
high-mass, rare halos, then overlap occurred inhomogeneously, and this
can also help to explain the observations.

In this paper we have planted the initial seeds of a very rich
phenomenological field. Observers and theorists should calculate and
compare the probability distribution of $\tau$ as a function of
redshift, and the correlation function of $\tau$ along the line of
sight. These statistical quantities will contain information on the
size distribution of \ion{H}{2} bubbles, on the neutral fraction
distribution within the bubbles, and on the density distribution of
neutral IGM outside the bubbles. This emerging field will benefit from
measurements of increasing resolution and signal to noise, made toward
many more lines of sight, and extended to higher redshifts. The
results in the near future should be a definitive detection of the
different stages of the reionization era, and the accumulation of a
large body of knowledge on the properties of ionizing sources and of
the IGM during reionization.

\acknowledgments

I thank Zoltan Haiman for useful discussions which motivated this
work, and George Djorgovski for helpful details regarding his
observations. I thank CITA for providing research support and a
stimulating research environment.


\begin{thebibliography}{}

\small

\bibitem[Abel \& Mo(1998)]{am98} Abel, T., \& Mo, H. J. 1998, ApJL, 
494, 151

\bibitem[Barkana \& Loeb(1999)]{me99} Barkana, R., \& Loeb, A. 1999, 
ApJ, 523, 54

\bibitem[Barkana \& Loeb(2001)]{me01} Barkana, R., \& Loeb, A. 2001,
Phys.\ Rep., 349, 125

\bibitem[Becker \etalb(2001)]{z6.3} Becker, R. H., Fan, X., White,
R. L., Strauss, M. A., \& Narayanan, V. K. 2001, AJ, submitted
(astro-ph/0108097)

\bibitem[Bond, Szalay, \& Silk(1988)]{bss88} Bond, J. R., Szalay, A. S., 
\& Silk, J. 1988, ApJ, 324, 627

\bibitem[Cen(2001)]{cen01} Cen, R.\ 2001, ApJ, in press (astro-ph/0101197)

\bibitem[Chiu \& Ostriker(2000)]{chiu} Chiu, W. A., \& Ostriker,
J. P. 2000, ApJ, 534, 507

\bibitem[Djorgovski \etalb(2001)]{z5.8} Djorgovski, S. G., Castro,
S. M., Stern, D., \& Mahabal A. 2001, ApJL, submitted (astro-ph/0108069)

\bibitem[Fan \etalb(2000)]{f00} Fan, X., \etal 2000, AJ, 120, 1167

\bibitem[Gnedin(2000)]{g00} Gnedin, N. Y. 2000, ApJ, 535, 530

\bibitem[Gnedin \& Ostriker(1997)]{go97} Gnedin, N. Y., \& Ostriker,
J. P. 1997, ApJ, 486, 581

\bibitem[Gunn \& Peterson(1965)]{GP} Gunn, J. E., \& Peterson, 
B. A. 1965, ApJ, 142, 1633

\bibitem[Haiman \etalb(2000)Haiman, Abel, \& Rees]{har00}  Haiman, Z., Abel, 
T., \& Rees, M. J. 2000, \apj, 534, 11

\bibitem[Haiman \& Loeb(1997)]{hl97} Haiman, Z., \& Loeb, A. 1997, ApJ,
483, 21

\bibitem[Haiman \& Loeb(1998)]{hl98} Haiman, Z., \& Loeb, A. 1998, ApJ,
503, 505

\bibitem[Haiman \etalb(1997)Haiman, Rees, \& Loeb]{hrl97}  Haiman, Z., 
Rees, M. J., \& Loeb, A. 1997, \apj, 476, 458; erratum -- 1997, ApJ,
484, 985

\bibitem[Halverson \etalb(2001)]{Dasy} Halverson, N. W., Leitch,
E. M., Pryke, C., Kovac, J., \etal 2001, ApJ, submitted
(astro-ph/0104489)

\bibitem[Kitayama \& Ikeuchi (2000)]{ki00} Kitayama, T., \& Ikeuchi, S. 2000, 
ApJ, 529, 615

\bibitem[Lacey \& Cole(1993)]{lc93} Lacey, C. G., 
\& Cole, S. M. 1993, MNRAS, 262, 627

\bibitem[Lee \etalb(2001)]{Maxima} Lee, A. T., Ade, P., Balbi, A.,
Bock, J., \etal 2001, submitted (astro-ph/0104459)

\bibitem[Loeb \& Barkana(2001)]{me01b} Loeb, A., \& Barkana, R. 2001, 
Annu.\ Rev.\ Astron.\ Astrophys., 39, 19

\bibitem[Magorrian \etalb(1998)]{m98} Magorrian, J., et al.\ 1998, AJ, 
115, 2285

\bibitem[Miralda-Escud\'e(1998)]{jordi98} Miralda-Escud\'e, J. 1998, 
ApJ, 501, 15

\bibitem[Miralda-Escud\'e \etalb(2000)Miralda-Escud\'e, Haehnelt, \&
Rees]{jordi} Miralda-Escud\'e, J., Haehnelt, M., \& Rees, M. J. 2000,
\apj, 530, 1

\bibitem[Navarro \& Steinmetz(1997)]{ns97} Navarro, J. F., \& Steinmetz, M. 
1997, ApJ, 478, 13

\bibitem[Netterfield \etalb(2001)]{Boomerang} Netterfield, C. B.,
Ade, P. A. R., Bock, J. J., Bond, J. R., \etal 2001, ApJ, submitted
(astro-ph/0104460)

\bibitem[Quinn \etalb(1996)Quinn, Katz, \& Efstathiou]{qke96} Quinn, T., 
Katz, N., \& Efstathiou, G. 1996, MNRAS 278, L49

\bibitem[Scalo(1998)]{scalo} Scalo, J. 1998, in ASP conference series
Vol 142, The Stellar Initial Mass Function, eds. G. Gilmore \&
D. Howell, p. 201 (San Francisco: ASP)

\bibitem[Thoul \& Weinberg(1996)]{tw96} Thoul, A. A., \& Weinberg, D. H. 
1996, ApJ, 465, 608

\bibitem[Weinberg \etalb(1997)Weinberg, Hernquist, \& Katz]{whk97} Weinberg, 
D. H., Hernquist, L., \& Katz, N. 1997, ApJ, 477, 8

\bibitem[Wood \& Loeb(2000)]{wl00} Wood, K., \& Loeb, A. 2000, ApJ, 545, 86

\end{thebibliography}
\end{document}